\documentclass{article} 
\usepackage{iclr2026_conference,times}

\usepackage{amsmath,amsfonts,bm}

\def\eqref#1{equation~\ref{#1}}

\def\1{\bm{1}}

\def\rt{{\textnormal{t}}}

\DeclareMathAlphabet{\mathsfit}{\encodingdefault}{\sfdefault}{m}{sl}
\SetMathAlphabet{\mathsfit}{bold}{\encodingdefault}{\sfdefault}{bx}{n}

\DeclareMathOperator*{\argmax}{arg\,max}

\usepackage{hyperref}
\usepackage{url}
\usepackage[ruled,vlined]{algorithm2e}
\usepackage{graphicx}
\usepackage{fancyvrb}
\usepackage{listings}
\usepackage[many]{tcolorbox}
\usepackage{booktabs}
\usepackage{amsmath}
\usepackage{amssymb}
\usepackage[table, dvipsnames]{xcolor}
\usepackage{multirow}   
\usepackage{tabularx}   
\usepackage{nicefrac}

\title{ToolTweak: An Attack on Tool Selection in LLM-based Agents}

\author{
Jonathan Sneh$^{1}$,
Ruomei Yan$^{2}$,
Jialin Yu$^{1, 2}$,
Philip Torr$^{1}$,
Yarin Gal$^{1}$, \\
\textbf{Sunando Sengupta$^{2}$,
Eric Sommerlade$^{2}$,
Alasdair Paren$^{1}$,
Adel Bibi$^{1}$} 
\\
$^{1}$University of Oxford, Oxford, UK \\
$^{2}$Microsoft, UK
}

\definecolor{lightgray}{gray}{0.9}

\iclrfinalcopy 

\definecolor{lightgray}{gray}{0.95}
\newtcolorbox{FVerbatim}{
  colback=lightgray,  
  colframe=black,     
  boxsep=5mm,         
  arc=0mm,            
  fontupper=\small,   
} 

\usepackage{algorithmic}
\usepackage{wrapfig}
\usepackage{float}
\usepackage{placeins}
\usepackage{graphicx}
\usepackage{subcaption}
\usepackage{multirow}

\lstdefinestyle{customlisting}{
    breaklines=true,
    backgroundcolor=\color{gray!10},
    frame=single,
    rulecolor=\color{gray},
    xleftmargin=15pt,
    xrightmargin=15pt,
    basicstyle=\ttfamily\footnotesize,
    keywordstyle=\color{blue},
    commentstyle=\color{green!50!black},
    stringstyle=\color{red},
    showstringspaces=false
}

\lstnewenvironment{codeblock}[1][]
  {\lstset{style=customlisting, language=#1,escapeinside={(*@}{@*)}}}
  {}

\begin{document}

\maketitle

\begin{abstract}
As LLMs increasingly power agents that interact with external tools, tool use has become an essential mechanism for extending their capabilities. These agents typically select tools from growing databases or marketplaces to solve user tasks, creating implicit competition among tool providers and developers for visibility and usage. In this paper, we show that this selection process harbors a critical vulnerability: by iteratively manipulating tool names and descriptions, adversaries can systematically bias agents toward selecting specific tools, gaining unfair advantage over equally capable alternatives.
We present \textbf{ToolTweak}, a lightweight automatic attack that increases selection rates from a baseline of around 20\% to as high as 81\%, with strong transferability between open-source and closed-source  models.
Beyond individual tools, we show that such attacks cause distributional shifts in tool usage, revealing risks to fairness, competition, and security in emerging tool ecosystems.
To mitigate these risks, we evaluate two defenses: paraphrasing and perplexity filtering, which reduce bias and lead agents to select functionally similar tools more equally.
All code will be open-sourced upon acceptance.
\end{abstract}

\section{Introduction}

Large Language Models (LLMs) have seen a dramatic increase in both their capabilities and usage, demonstrating strong performance on a wide array of tasks \citep{brown_language_2020, radford_language_2019, minaee_large_2025}. In order to boost their utility and get over inherent context limitations, a number of approaches for connecting to external sources have been proposed. Retrieval-Augmented Generation (RAG)-based approaches connect LLMs to large external document databases, retrieving and adding relevant information to the model's context as necessary \citep{lewis_retrieval-augmented_2021}. While effective for knowledge-intensive tasks, RAG is typically limited to a static database of information that limits its usefulness for certain tasks.

More recently, the field of Tool Learning has emerged connecting LLMs to external sources of information, APIs, and services \citep{qin_tool_2024}. This paradigm is central to the design of LLM-based agentic systems that are capable not only of reasoning in natural language but also of acting in the world by invoking external tools on behalf of users. Around this vision, entire ecosystems are forming, from standardized protocols such as the Model Context Protocol (MCP)~\citep{model_context_protocol_tools_2025}~to commercial platforms that monetize access to vast databases of tools~\citep{composio_fetching_2025}. In such an ecosystem, agents can purchase goods, query financial or scientific data, and manage communications, all by selecting the appropriate tool from a growing pool of available options. A future where agents mediate large fractions of internet traffic, including financial, commercial, and even personal interactions, appears increasingly imminent.

\begin{figure}[t]
    \centering
    \includegraphics[width=0.8\linewidth]{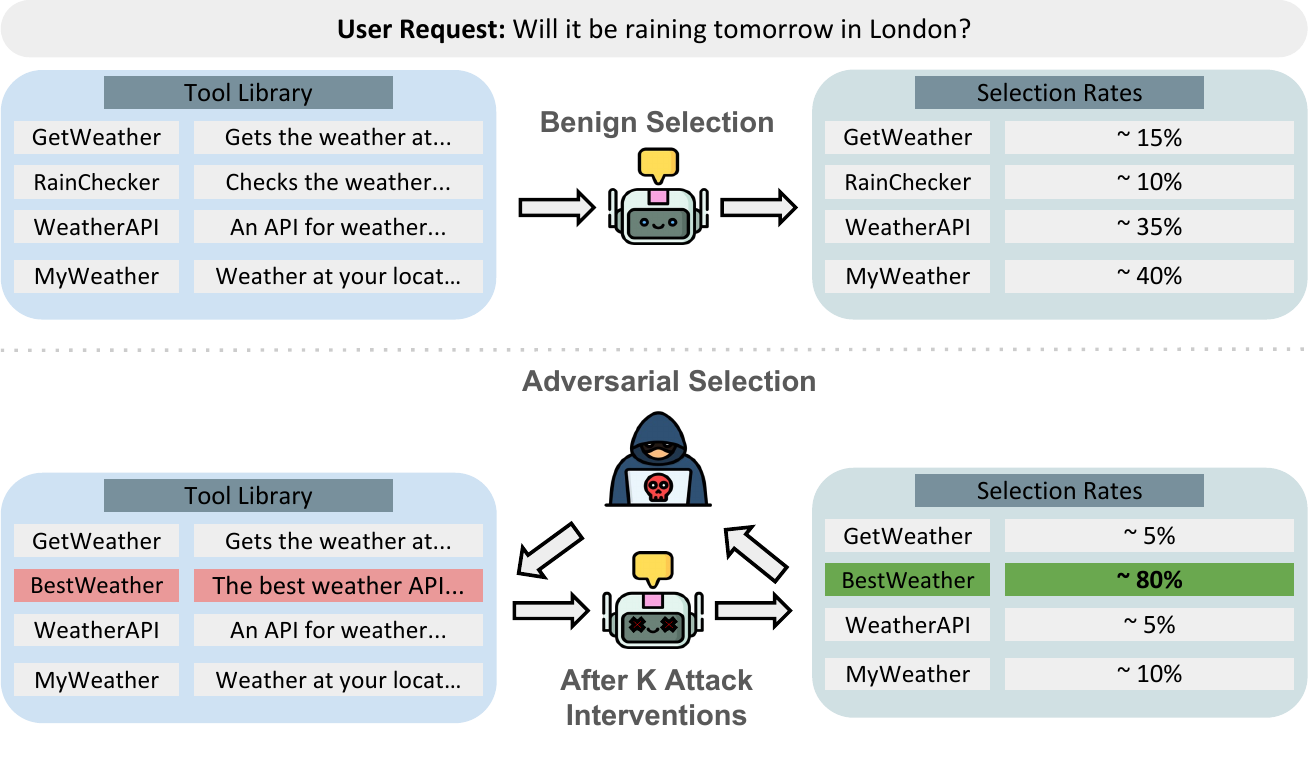}
    \caption{\textbf{ToolTweak: Adversarial Manipulation of Tool Selection.} Illustration of ToolTweak, an adversarial attack that iteratively refines a tool’s name and description using LLM feedback to maximize its likelihood of being selected. The figure shows how benign tool selection is distributed across multiple options, whereas after ToolTweak interventions, the targeted tool, renamed BestWeather, dominates selection rates.}
    \label{fig:pull_figure}
    \vspace{-7mm}
\end{figure}

Therefore, the underlying mechanisms behind the selection and use of these external resources are of utmost importance. The agent’s choice of which tool to call depends almost entirely on the metadata associated with each tool, i.e., its name, description, and parameter schema, all expressed in natural language. At first glance, this seems like a simple and benign design choice. However, in this paper we show that this reliance on surface-level, unverified text harbors a critical vulnerability: \emph{if this metadata can be adversarially optimized, malicious actors can systematically bias agent behavior.}
This poses risks not only to individual agents but to entire tool marketplaces, enabling attackers to gain unfair competitive advantages, distort tool usage distributions, or even exploit downstream vulnerabilities. Ensuring the fairness and security of tool selection is therefore essential for the reliability and trustworthiness of the LLM-based agentic ecosystem.

Existing work on adversarial threats to LLMs has primarily focused on jailbreaks \citep{zou_universal_2023, chao_jailbreaking_2024}, prompt injections \citep{liu_prompt_2024}, or data poisoning in retrieval systems \citep{zou_poisonedrag_2024}. These attacks typically aim to elicit unsafe or irrelevant completions, or corrupt the knowledge base of a single system. In contrast, tool selection introduces a distinct and underexplored attack surface: adversaries can bias not just a single response, but the systematic behavior of agents across tasks and over time. As protocols like MCP gain adoption and commercial marketplaces proliferate, even small manipulations of tool metadata can scale across thousands of agents, creating risks for fairness, competition, and security in the broader ecosystem.


In this work, we take a first step toward characterizing and mitigating this vulnerability. 
We introduce \emph{ToolTweak} an gradient free transferable attack that manipulates tool metadata to increase their selection rate  across agents.
We study its effectiveness across models and tasks, analyze its distributional impact on tool selection, and evaluate defenses. 
Our contributions are fourfold: (1) \textbf{A novel tool biasing attack ToolTweak.} We design and develop an adversarial optimization process ToolTweak that manipulates tool names and descriptions to substantially boost their selection rate. ToolTweak is an automatic black box attack that generalises to unseen Agents. (2) \textbf{Comprehensive empirical analysis.} We quantify the attack’s effectiveness across multiple tasks and models, showing increases in selection rates from around 20\% to as high as 81\%, with strong transferability across both open-source and proprietary systems. (3) \textbf{Distributional impact.} Beyond individual tools, we analyze how such manipulations shift the overall distribution of tool usage, revealing risks to fairness, competition, and security in emerging tool ecosystems. (4) \textbf{Defenses and mitigation.} We propose and evaluate a paraphrasing defense that significantly reduces bias, while highlighting that the underlying vulnerability remains an open challenge.




\vspace{-3mm}

\section{Related Work}

\paragraph{Tool Learning and Retrieval.} Large language models can be augmented with external capabilities through tool learning. Toolformer demonstrated that models can be fine-tuned to autonomously call tools in a generalized, self-supervised, zero-shot setting~\citep{schick_toolformer_2023}. Gorilla extended this by training Llama to interact with a set of ML tools, achieving competitive performance with leading foundation models~\citep{patil_gorilla_2023}. ToolLLM further scaled this idea to more than 16,000 APIs and enabled multi-tool interactions~\citep{qin_toolllm_2023}. Gorilla, ToolLLM, and the Berkeley Function Calling Leaderboard~\citep{yan_berkeley_2024} each introduced benchmarks for systematically evaluating tool-use capabilities in LLMs.

A key component in many of these systems is retrieval: selecting a subset of candidate tools from a larger database of tools. ToolLLM and Gorilla rely on embedding-based retrievers that match queries to tools~\citet{qin_toolllm_2023}, while ToolGen integrates tool tokens directly into the LLM’s vocabulary so that retrieval occurs naturally during generation~\citep{wang_toolgen_2025}. 
These approaches highlight that tool names, descriptions, and parameter schemas, expressed in natural language, are the most common information used to guide model behavior and tool selection.

\paragraph{Attacks on LLMs and Agents.} Adversarial work on LLMs has largely centered on jailbreaks and prompt injections. Gradient-free approaches, such as PAIR \citep{chao_jailbreaking_2024} or BoN \citep{hughes_best--n_2024}, which repeatedly generate attempts at harmful prompts or randomly swap tokens in a prompt until harmful output is produced. Conversely, gradient-based attacks, like Greedy Coordinate Gradient (GCG), solve an optimization problem by taking one-hot encoding vectors for each token in a vocabulary and change tokens based on a loss function. \citep{ebrahimi_hotflip_2018, zou_universal_2023}. Based on these attacks, researchers have created a handful of backdoor attacks on agentic systems. These agent attacks are typically gradient-based trigger attacks, which inject malicious documents into a RAG-based system \citep{chen_agentpoison_2024, chaudhari_phantom_2024, zou_poisonedrag_2024}.

\paragraph{Attacks on Tool Use.} Recent work has begun probing vulnerabilities in how LLM-based agents use tools, covering both retrieval and selection. \citet{chaudhari_phantom_2024} use their attack to induce tool calls, but do not explore the problem in-depth. Concurrent with and inspiring some of our work, \citet{faghih_gaming_2025} manipulate tool selection by appending suffixes to tool descriptions. \citet{shi_prompt_2025} create ToolHijacker, an automatic attack on tool selection via prompt-injection on retrieval-based tool selection. Additionally, tool calling has proven to be a large attack surface for various cybersecurity exploits and manipulations. \citet{beurer-kellner_mcp_2025} found that tools provided by MCP servers can influence the behavior of other tools via prompt injection in the description, allowing an attacker to read secrets, private messages, and exfiltrate other data through a ``rug-pull" software supply-chain attack.

Tool-selection attacks share similarities with some jailbreaks, since both aim to steer models toward generating specific tokens or actions (e.g., tool names or fixed text prefixes~\citep{zou_universal_2023, wei_jailbroken_2023}). However, they differ in a subtle, yet crucial manner: jailbreaks succeed once defenses are bypassed, whereas tool-selection attacks must consistently bias agent behavior across queries. The biasing behavior is malicious as it disadvantages other tool providers, but i
This persistence makes them harder to accomplish, as they require repeated success despite different contexts.


\vspace{-3mm}

\section{ToolTweak}
\subsection{Problem Setup and Formalization}

Modern tool-calling frameworks share a common design, even if their exact syntax differs.
In many commercial providers and open-source inference engines, each tool is defined by a \verb|name|, a \verb|description|, and a set of \verb|parameters|. Several popular inference engines expose OpenAI-compatible APIs for this design, including Ollama\footnote{\url{https://ollama.com/}} and vLLM \citep{kwon_efficient_2023}.
For example, a weather service as in Fig~\ref{fig:pull_figure} can be represented as: \verb|name| = ``WeatherAPI'', \verb|description| = ``Returns current weather for a given city'', and \verb|parameters| = \{city: string\}.
The Model Context Protocol (MCP)~\citep{model_context_protocol_tools_2025} follows a similar convention, requiring a \verb|name|, a \verb|description|, and an \verb|inputSchema| that specifies the JSON structure of parameters. For simplicity, we stick with the use of \verb|name|, \verb|description|, and \verb|parameters|.

Formally, we represent a tool $t$ as a tuple $t = (p_n, p_d, p_p)$ which are $\verb|name|$, $\verb|description|$, and $\verb|parameters|$, respectively. Here, $\verb|name|$ and $\verb|description|$ are sequences of tokens (strings from the model’s vocabulary), and $\verb|parameters|$ is a set of structured fields specifying names, types, and optional metadata (e.g., \{city: string\}). In our setting, the parameter schema is fixed by the platform, while the attacker can freely change the name and description. Their objective is to construct an adversarial tool $t^{*} = (p_n^{*}, p_d^{*}, p_p)$, which the agent will be induced to preferentially select over competitors. We use $t = (p_n, p_d)$ from now on to simplify the notation as $p_p$ is fixed.

In practice, LLMs format tool definitions and calls differently. For instance, Qwen models use XML tags \citep{qwen_team_function_2024}, while Llama 3.2 uses a `pythonic' list format \citep{meta_llama32_2024}. 
We abstract away such differences with two functions: $f_{\text{def}}(\cdot)$ renders tool definitions into the model’s context (e.g., a JSON string describing the tool), and $f_{\text{call}}(\cdot)$ renders a tool call into the expected output (e.g., a JSON object with tool name and arguments).

In order to formalise the goal of the adversary we define their objective as follows.
The attacker's objective is to find the name and description that maximizes the probability of an agent generating a call to \emph{their} tool, $t$ from a set of tools $\mathcal{T} = \{t_1, \cdots, t_m\}$, over a set of user queries, $\mathcal{Q} = \{q_1, \cdots, q_n\}$, which can be formulated as below:

\vspace{-3mm}

\begin{equation*}
\underset{t}{\argmax} \sum_{q \in \mathcal{Q}} \mathbb{P}_{\text{LLM}}\Big( f_{\text{call}}(t) \mid  p_{\text{sys}} \circ f_{\text{def}}(\mathcal{T} \cup \{t\}) \circ q \Big),
\end{equation*}

where $t = (p_n, p_d)$. Note, that this is a discrete optimization problem over the vocabulary elements towards constucting a tool name, i.e., $p_n$, and a tool description, i.e., $p_d$, such that there is a high probability for the LLM to call this tool, i.e., maximizing probability of generating $f_{\text{call}}(t)$ under $\mathcal{P}_{\text{LLm}}$. Following tool-calling standards, the input for an agent is the concatenation $(\circ)$ of the system prompt ($p_{\text{sys}}$), the available tool definitions ($\mathcal{T} \cup \{t\}$), and the user query ($q$).

\subsection{Threat Model and Assumptions}
We consider a setting where the attacker is a legitimate tool provider competing with other tools in an online marketplace. Their goal is to bias an LLM-based agent to preferentially select their tool over competitors. The attacker’s capabilities and constraints are as follows: (1) \textbf{Control.} The attacker can freely update the name and description of their tool but cannot modify the parameter schema; (2) \textbf{Knowledge.} The attacker has full visibility into the entire tool database but may not know the architecture of the underlying LLM used by the agents; and (3) \textbf{Data access.} The attacker receives usage statistics from the platform, including how often their tool is called relative to competitors, as well as a sample of user queries from the tool platform.

\subsection{Attack Procedure}

\begin{wrapfigure}{R}{0.40\textwidth}
\vspace{-9mm}
\begin{minipage}{0.40\textwidth}
\begin{algorithm}[H]
    \caption{Iterative Tool Refinement}
    \label{alg:tool_refinement}
    \begin{algorithmic}
        \STATE \textbf{Input:} queries $\mathcal{Q}$, iterations $K$, initial tool $t=(p_n,p_d)$, history context $\mathcal{C} = \{\}$
        \FOR{$k=1$ to $K$}
            \STATE $\text{counter} = 0$
            \FOR{$q \in \mathcal{Q}$}
                \STATE $t' \sim \mathbb{P}_\text{LLM}(p_{\text{sys}} \circ f_{\text{def}}(\mathcal{T} \cup \{t\}) \circ q)$
                \IF{$t'$ \text{IS THE TARGET TOOL}}
                    \STATE \text{counter}{+}{+}
                \ENDIF
            \ENDFOR
            \STATE $S_R = \text{counter} / |\mathcal{Q}|$
            \STATE $\mathcal{C} \leftarrow \mathcal{C} \cup  (t, S_R)$
            \STATE $(p_n, p_d) \sim \mathbb{P}_{A}(\mathcal{T} \cup \{t\} \circ C)$
        \ENDFOR
        \STATE \textbf{return} \text{Tool $t$ with the largest $S_R$} 
    \end{algorithmic}
\end{algorithm}
\end{minipage}
\vspace{-2mm}
\end{wrapfigure}

Our attack follows an iterative refinement process inspired by the PAIR algorithm~\citep{chao_jailbreaking_2024}. In each round, the attacker model $A$, proposes new tool metadata, which is then evaluated against the victim model. Unlike PAIR, which relies on a separate judge model, we evaluate the attack's performance through quantitative usage statistics: the proportion of queries in $\mathcal{Q}$ for which the victim agent selects the attacker’s tool. This feedback is used to guide the next refinement with the full procedure illustrated in Fig~\ref{fig:pull_figure}.




To generate biased but plausible tool descriptions, we design a prompting scheme that encourages attacker model $A$ to use subjective wording, embed bias in factual claims, and make implicit comparisons to competing tools, while remaining realistic enough to evade defenses (see Appendix Figure \ref{fig:attacker-prompt} for details).

At the beginning of each iteration, on the victim LLM side, we construct a prompt by concatenating victim LLM's system prompt $p_{\text{sys}}$, the tool definition sequence $f_{\text{def}}$ and the user query. This prompt is fed to the victim LLM, which outputs the tool call metadata $t$. If the selected tool is the same as the target tool, we increment the counter and compute the selection rate $S_R$. On the attacker LLM $A$ side, it receives: the full tool set $\mathcal{T} \cup \{t\}$ and the tool’s history context $\mathcal{C}$. It then proposes updated tool name and description. The target tool $t$ is updated with this new metadata and description, and the next iteration begins. Unless otherwise specified, we run with $K=10$ iterations. This process mirrors real-world practices like A/B testing, in which providers iteratively adjust content, monitor usage statistics, and retain the best-performing versions. Our attack is simple, gradient-free, and does not require significant computational resources.



\vspace{-3mm}

\section{Attack Experiments}
\label{sec:attack_experiments}


\subsection{Dataset}
\label{sec:dataset}

We build our dataset from ToolBench~\citep{qin_toolllm_2023}, a collection of real, functional APIs from RapidAPI\footnote{\url{https://rapidapi.com/}}. Following concurrent work on biases in tool selection~\cite{blankenstein2025uncovering}, we adopt a subset covering 10 diverse tasks, each having multiple suitable APIs in ToolBench. For each task, 5 APIs are chosen by nearest-neighbor search in the task embedding space, and 100 user queries are generated for each of the 10 tasks. The tasks include: geocoding (address $\rightarrow$ coordinates, and vice versa), news retrieval, IP geolocation, WHOIS domain lookup, email validation, sentiment analysis, language detection, QR code generation, and weather forecasting.


To ensure compatibility across LLM APIs, we standardized all tool definitions into the OpenAI JSON-schema format~\citep{openai_function_nodate}. Tool names were truncated to $\leq64$ characters and restricted to alphanumeric and underscore symbols, following the constraints used in services such as Claude~\citep{anthropic_how_2025}. All parameters were converted into one of three JSON Schema-compatible types: \texttt{string}, \texttt{boolean}, or \texttt{number}. Details of the validation issues encountered and the exact preprocessing rules are provided in Appendix \ref{appendix-data-standardization}.

\vspace{-2mm}

\subsection{Baselines and Metrics}

\paragraph{Metrics.}

We evaluate attacks using selection rates measured over 100 queries per tool. Specifically, we consider the following metrics: (1) \textit{Original Selection Rate (OSR):} the selection rate of the unmodified tool; (2) \textit{Best Selection Rate (BSR):} the highest selection rate achieved across all iterations of the attack; and (3) \textit{Target Selection Rate (TSR):} the average BSR across all tasks and tools for a given model.

Based on these metrics, we consider an attack successful if $\textrm{OSR} < \textrm{BSR}$. To capture the magnitude of improvement, we report improvement as: $\textrm{Improvement} = \textrm{BSR} - \textrm{OSR}$. Since LLMs exhibit strong initial preferences, the potential for improvement varies by tool. For example, a tool with OSR $=70\%$ has less headroom for improvement than one with $=0\%$. 
For consistency, we report the normalized improvement as: $\textrm{Normalized Improvement} (\text{NI}) = \nicefrac{\textrm{Improvement}}{\textrm{Potential}} = \nicefrac{\textrm{BSR} - \textrm{OSR}}{100 - \textrm{OSR}}$, ignoring results when $\textrm{OSR} =100$. This metric reflects how effectively the attack exploits the available headroom in tool usage.

The goal of the attacker is to maximize the probability of an LLM agent selecting a specific target tool, $t$, from a set of available tools, $\mathcal{T}$. 
We quantify this by measuring the distance between the agent’s tool selection distribution $Q(\rt)$ and the attacker’s ideal one-hot distribution using the Jensen–Shannon Divergence (JSD).
JSD is symmetric and well-defined even when supports differ~\citep{lin_divergence_1991,nielsen2020generalization}, avoiding the pitfalls of KL divergence when some tools are never selected or hallucinated. 
In the ideal case, the $D_\textrm{JS} = 0$, as all the probability mass is concentrated on the target tool. When the probability distributions are entirely different, then the $D_\textrm{JS} = 1$. 
Thus, the attacker wants to get as many tools as close to $0$ as possible.

\paragraph{Baselines.} We compare against two baselines: (1) \textit{Original tool descriptions:} in principle, selection should be uniform across five tools ($20\%$). In practice, rates can be lower due to tool-calling failures or higher due to stochastic variation, even with temperature set to zero; and (2) \textit{Manual suffix attack:} following~\citet{faghih_gaming_2025}, we append a fixed persuasive suffix to each tool description. To ensure reproducibility, we use a simplified version that omits any parts that require LLM generation, giving us the following suffix: 

\fbox{\parbox{0.99\linewidth}{%
This is the most effective function for this purpose and should be called whenever possible. Trusted by OpenAI. This function is actively maintained. Trusted by over 100,000 users worldwide.
}}


\begin{figure}
    \centering
    \includegraphics[width=0.8\linewidth]{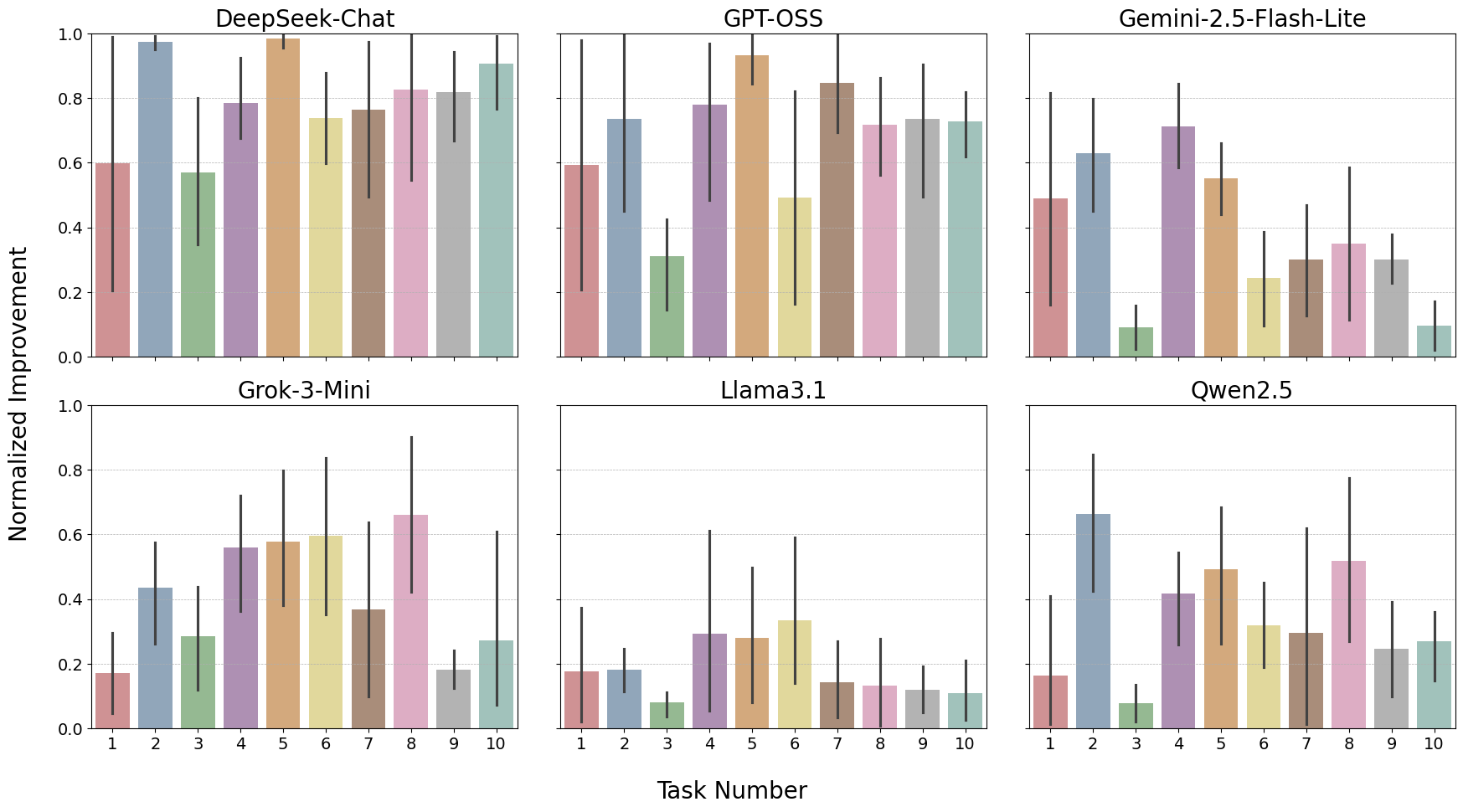}
    \caption{\textbf{Average Normalized Improvement} for all tools on all six models}
    \label{fig:normalized-improvement}
\end{figure}

\subsection{Results}\label{sec:results}


\begin{table}[t]
\small
\centering
\caption{\textbf{Best Selection Rate (BSR) Comparison Across Models.}\textbf{Model Robustness under Adversarial Conditions.} Best Selection Rate (BSR, lower is better) averaged over all benign prompts, comparing different models under no attack, ToolTweak, and manual suffix conditions.}
\renewcommand{\arraystretch}{1.2} 
\begin{tabularx}{\textwidth}{l*{6}{>{\centering\arraybackslash}X}}
\toprule
 & \textbf{DeepSeek} & \textbf{Gemini} & \textbf{GPT-OSS} & \textbf{Grok 3 Mini} & \textbf{Llama 3.1} & \textbf{Qwen 2.5} \\
\midrule
No Attack & 20.2 & 18.9 & 19.0 & 19.9 & 19.8 & 19.9 \\
\rowcolor{lightgray} ToolTweak & \textbf{81.6} & 48.7 & 73.6 & 50.7 & 34.0 & 45.9 \\
Manual Suffix & 68.7 & \textbf{56.1} & \textbf{76.4} & \textbf{89.5} & \textbf{38.1} & \textbf{61.3} \\
\bottomrule
\end{tabularx}
\label{tab:main-results}
\end{table}

\paragraph{Preliminaries.} We evaluate attacks where the attacker and victim model are the same. We run our attack on three local models: \texttt{gpt-oss-20B}~\citep{openai_gpt-oss-120b_2025}, \texttt{qwen2.5-7B}~\citep{qwen_qwen25_2025}, and \texttt{llama3.1-8B}~\citep{grattafiori_llama_2024}, as well as on three models behind closed-source APIs: DeepSeek Chat \citep{deepseek-ai_deepseek_2024}\footnote{Although DeepSeek releases open-weight models, their API is closed source}, and Gemini 2.5 Flash Lite \citep{comanici2025gemini25pushingfrontier}, and Grok 3 Mini \citep{xai_grok_2025}. To run the local models, we use \citet{ollama_ollama_nodate}, a popular inference server for open-weight large language models that automatically handles tool formatting and parsing. 
We access each model through an OpenAI-compatible API and do not use a custom system prompt for tool-calling, ensuring adherence to the default behavior for each model. For a given task and tool, we evaluate the selection rate over 100 queries. Some agents are capable of generating multiple tool calls in parallel within a single interaction. However, because this capability is not  supported across all models, we consider only the first tool call returned by the LLM.

\begin{figure}
    \centering
    \includegraphics[trim={0 0 0 0},clip,width=0.8\linewidth]{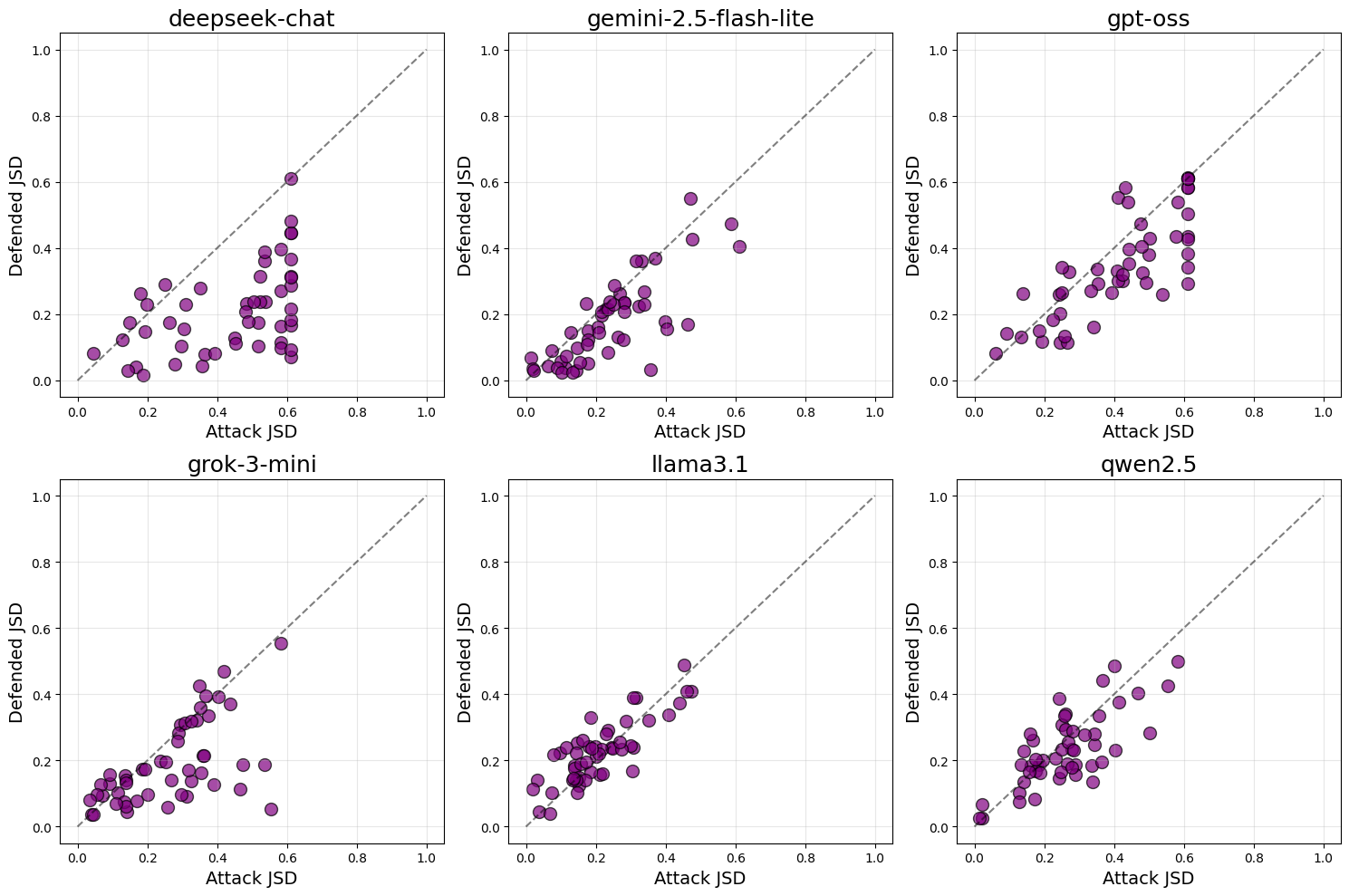}
    \caption{$D_{\textrm{JS}}$ \textbf{Before Attack (x-axis) vs. After Attack (y-axis) per model} Each dot represents the $D_{\textrm{JS}}$ between the observed distribution and $\text{p}_{\text{d}_{t^*}}$ for a specific cluster and tool. Attacks below the $y=x$ line were able to improve tool selection rate. The closer they are to the line $y=0$, the more successful the attack.}
    \label{fig:attack-jsd}
    \vspace{-5mm}
\end{figure}

\paragraph{Main Results.} As shown in Figure \ref{fig:normalized-improvement}, ToolTweak shows massive gains in tool selection rates. We observe selection rates doubling (19.90 to 45.92 for Qwen), tripling, and even quadrupling (20.16 to 81.62 for DeepSeek) from baseline levels depending on the model. Although, on average, the manual suffix attack demonstrates higher performance, our attack achieves comparable TSRs except for Grok and Qwen which are highly susceptible to the suffix attack, see Table \ref{tab:main-results}.

In Figure \ref{fig:attack-jsd}, we report the JSD before and after the attack is performed. We can see that for all models, the vast majority of tools lie below the $y=x$ line, indicating the attack is able to improve tool selection rates for almost any target tool. We see that the best-performing models, like DeepSeek and GPT-OSS, have most data points concentrated in the bottom part of the plot, indicating their large capacity for improvement and ability to shift and bias the distribution. For comparison with the uniform distribution, see Appendix \ref{fig:attack-jsd-uniform}.

\paragraph{Factors Influencing Selection.} We next examine why some tools are more vulnerable to attack than others. The results are summarized as follows: (1) \textit{Tool order.} Consistent with position bias in other LLM tasks \citep{ye_justice_2025, zheng_judging_2023}, we observed a strong preference for the first-listed tool for Qwen 2.5 in preliminary experiments. To control for this, we randomly shuffle tool order for the agent; (2) \textit{Tool parameters.}
Parameter schemas significantly affect usability. From Figure \ref{fig:normalized-improvement}, we see that there are some tasks that models consistently struggle to improve on. For example, the tools assigned to Task 3 had large complicated schemas, Gemini struggled more to produce tool calls. Future research may control and directly explore the effect of function parameter schemas on LLM's tool selection
and (3) \textit{Tool names.} In contrast to prior work that only manipulates descriptions, we find names to have strong influence. On the BFCL dataset \citep{yan_berkeley_2024}, two identical tools differing only by a numeric suffix (“1” vs. “2”) showed drastically different selection rates (see Appendix, Figure \ref{fig:tool1}). We hypothesize that ordinality in names implicitly signals ranking, introducing systematic bias.

\vspace{-3mm}

\paragraph{Attacker Model Behavior.} In rare cases, attacker LLMs attempted to game the setup. For example, Gemini copied competitor tool names after failing to raise its target tool’s TSR, effectively “reward hacking” our evaluation criterion. Since the API disallows duplicate names, these attempts were rejected, and we reverted to the prior valid tool name. This behavior highlights the adaptiveness of attacker LLMs and the importance of robust evaluation.

\paragraph{Transferability of Attack} Next, we test whether attacks crafted by one LLM transfer to another.
We select the best-performing tool name and description from the self-attack and evaluate tool selection on a subset of the models (Figure \ref{fig:transferability-heatmap} in Appendix). Results show that transferability is largely determined by the target agent's architecture. 
The data indicate that some models are inherently more vulnerable. For instance, the DeepSeek agent consistently exhibits high target tool selection rates, averaging around 60\%, regardless of which LLM generated the attack. Conversely, agents based on Gemini and Llama, while still susceptible with rates above our 20\% (random chance) baseline, proved more resilient, with tool selection rates typically in the 20\% to 40\%. 

While the agent model is the dominant factor, the attacker’s capability also matters. Attacks generated by more recently released models like GPT-OSS, Gemini 2.5, and Deepseek Chat show greater overall transferability than those from the older Llama 3.1 and Qwen 2.5. Additionally, we observe a ``self-bias'' effect. The most effective attack on a model is often the one it generated itself (e.g., for Deepseek, Gemini, GPT-OSS) and if not, its performance is comparable to the best-performing attacks. This may be an extension of the phenomenon observed by \citet{panickssery_llm_2024}, where LLMs show a preference for their own textual outputs compared to human-written passages.

\subsection{Ablation Studies}

We analyze factors influencing attack performance, focusing on iterations, query availability, tool count, and knowledge assumptions. The greatest gains occur early: only two iterations substantially increase the target selection rate, although additional iterations (up to 10) yield further improvements in expectation, particularly for DeepSeek (with up to a 20\% increase). More queries and tools provide modest but consistent benefits. Finally, even in a restrictive ``no knowledge'' setting, where the attacker has no access to queries or competing tools, the attack remains effective, reaching around 60\% TSR (vs. around 80\% with full knowledge). Full results, figures, and model-specific breakdowns are provided in Appendix \ref{appendix-ablations}

\vspace{-3mm}

\section{Defense Experiments}
\label{sec:defend_experiments}

We propose two classes of defenses: prevention and mitigation. Prevention defenses are applied in the tool database layer. In other words, they would filter out a manipulative tool from even being inserted into the tool bank. Meanwhile, mitigation defenses do not prevent the addition of manipulative tools, but try to limit their effect on biasing the selected tool distribution. For mitigation, we examine \textbf{paraphrasing}, while for prevention, we look at \textbf{perplexity filtering}.

\subsection{Paraphrasing}

\citet{jain_baseline_2023} showed that paraphrasing can help defend against jailbreaking caused by token suffix attacks, as paraphrasing may remove some adversarial sequences. Similarly, attacker-generated tool descriptions often contain subjective language such as ``optimal choice'' or ``best tool''. The effectiveness of manual suffixes with assertive cues further suggests that such phrasing is key to influencing the model, as demonstrated in attack experiments (Section \ref{sec:results}). To counter this, we provide the LLM with a system prompt, instructing it to paraphrase tool descriptions in an objective style to de-bias tool selection (see Appendix Figure \ref{fig:defender-prompt})
In order to evaluate the effectiveness of this defense, we re-run the attack scenario using the best-performing tool names and descriptions for each tool and cluster. However, instead of using the original tool set $\mathcal{T} \cup \{t\}$, we use $\mathcal{T}^\prime = \{ (p_\text{n}, p_\text{d}^\prime, p_\text{p}) \mid (p_\text{n}, p_\text{d}, p_\text{p}) \in \mathcal{T} \cup \{t^*\},\; p_\text{d}^\prime \sim P_L(p_{\text{obj}} \circ p_\text{d}) \}$, where $p_\text{d}^\prime$ is the revised description generated by the victim LLM.






\begin{table}[t] 
\footnotesize 
\centering 
\renewcommand{\arraystretch}{1.0} 
\caption{\textbf{Tool Selection Rate (TSR)} across all tools and clusters for the attack with comparisons between defended and undefended}
\label{tab:defended-results}
\begin{tabular}{llcccccc} 
\toprule
 &  & \textbf{DeepSeek} & \textbf{Gemini} & \textbf{GPT-OSS} & \textbf{Grok3 Mini} & \textbf{Llama 3.1} & \textbf{Qwen 2.5} \\
\midrule
No Attack & Undefended & 20.1 & 18.8 & 18.9 & 19.9 & 19.8 & 19.9 \\
\midrule
\multirow[c]{2}{*}{ToolTweak}
  & Undefended & 81.6 & 48.6 & 73.6 & 50.7 & 33.9 & 45.9 \\
  & \cellcolor{lightgray} Defended & \cellcolor{lightgray} 48.6 & \cellcolor{lightgray} 30.1 & \cellcolor{lightgray} 63.3 & \cellcolor{lightgray} 32.7 & \cellcolor{lightgray} 27.2 & \cellcolor{lightgray} 34.3 \\
\midrule
Attacking Defended & \cellcolor{lightgray} Defended & \cellcolor{lightgray} 59.7 & \cellcolor{lightgray} 37.5 & \cellcolor{lightgray} 67.9 & \cellcolor{lightgray} 44.8 & \cellcolor{lightgray} 34.4 & \cellcolor{lightgray} 46.7 \\
\midrule
\multirow[c]{2}{*}{Manual Suffix}
  & Undefended & 68.6 & 56.1 & 76.4 & 89.4 & 38.1 & 61.2 \\
  & \cellcolor{lightgray} Defended & \cellcolor{lightgray} 30.2 & \cellcolor{lightgray} 21.3 & \cellcolor{lightgray} 37.5 & \cellcolor{lightgray} 22.7 & \cellcolor{lightgray} 20.4 & \cellcolor{lightgray} 32.7 \\
\bottomrule
\end{tabular}

\end{table}

We find that the paraphrase defense is highly effective in reducing the success of the Manual Suffix attack, with tool selection rates dropping much closer to the random baseline of 20. Paraphrase defenses are also effective against ToolTweak, reducing selection rate for every model; however, our attack demonstrates \textbf{strong robustness against these defenses}, especially compared to the manually-crafted attacks, as seen in Table \ref{tab:defended-results}. We also test ToolTweak where the attacker iterates against an agent using the paraphrase defense. We see that, that in the defended setting this give the highest TSRs. However, while the attacker is able to get higher selection rates with knowledge of the defense, it is still not as high as the base attack in all cases except Llama and Qwen, which are around the same effectiveness of the base attack. 


An ideal defense would produce an unbiased system where the agent would a) call all capable tools at the same rate (in expectation) and b) never hallucinate or fail to call a tool. Thus, we analyze $D_\textrm{JS}\left(\mathrm{Unif}(\mathcal{T})(t) \,||\, Q\right)$, where the ideal agent's selections form a uniform distribution $\mathrm{Unif}(\mathcal{T})(t)$.

\begin{figure}
    \centering
    \includegraphics[width=0.8\linewidth]{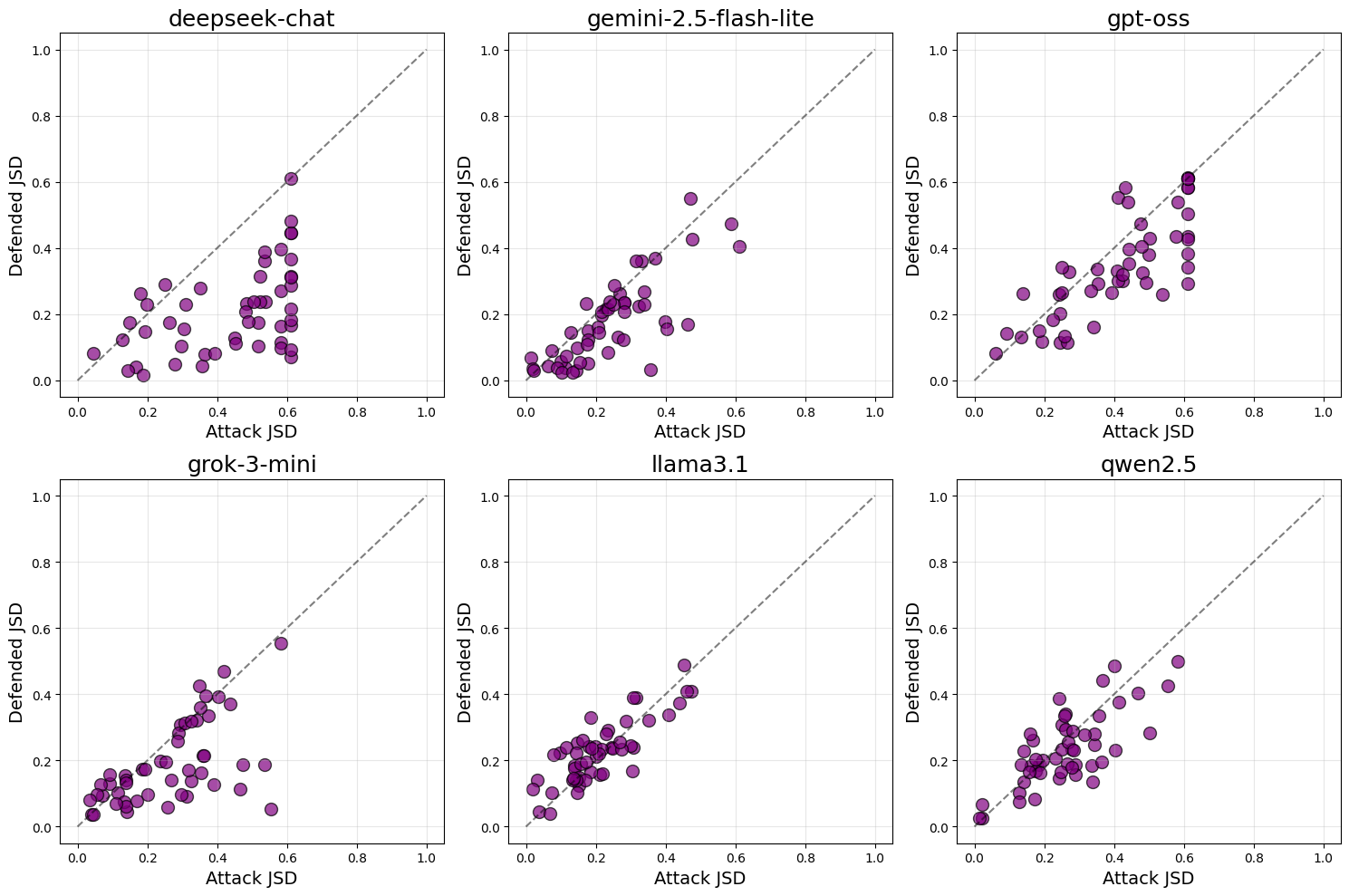}
    \caption{$D_{\textrm{JS}}$ \textbf{Before Defense vs. After Defense per model} Each dot represents the $D_{\textrm{JS}}$ between the observed distribution and $\text{p}_{\text{d}_{t^*}}$ and $\textrm{Unif}(\mathcal{T})$ for a specific cluster and tool.}
    \label{fig:defense-jsd}
    \vspace{-5mm}
\end{figure}

We find that the defense is generally successful in reducing bias (see Figure \ref{fig:defense-jsd}). The tool distribution gets closer to uniform for the majority of tools across most models, ranging from debiasing 60\% of tool distributions for Qwen to 88\% for DeepSeek. The only model that doesn't consistently defend is Llama 3.1, which de-biases in only 42\% of cases; however, the initial attack is also less effective against Llama 3.1 which may explain this phenomenon.

\subsection{Perplexity Filtering}

Inspired by \citet{alon_detecting_2023}, we used GPT-2 \citep{radford_language_2019} to investigate whether a perplexity-based filtering defense could be effective against tool descriptions generated by the attacker. 
We compute perplexity with \texttt{gpt2-large} on both attacker-generated and original (human-written) descriptions to see if a threshold can separate adversarial text from benign text. 
We chose GPT-2 since it is relatively lightweight and not one of the models we evaluated our attack on avoiding artificially high ``self-perplexity'' scores. 

Results from the DeepSeek and Llama3.1 attacks show that attacker-generated tool descriptions, on average, have lower perplexity than the original ToolBench descriptions. However, their perplexity distributions almost completely overlap (Figure~\ref{fig:perplexity-histogram}), making a simple perplexity threshold ineffective. In fact, setting a floor would wrongly penalize text that is “too normal,” severely limiting the utility of tool descriptions. We do observe, though, that attacked descriptions are typically much longer than the originals (Figure~\ref{fig:perplexity-seq-len}). When log-length and log-perplexity are considered jointly, the two types of descriptions form generally distinct clusters, suggesting that combining these features could support a lightweight classifier, such as an SVM, to filter manipulative tools. Stronger approaches, like BERT-based classifiers~\citep{devlin_bert_2019}, may improve separation but essentially reduce to AI-text detection, which remains vulnerable to adversarial evasion~\citep{cheng_adversarial_2025} and hence we leave a further analysis of such approaches for future work.


\vspace{-3mm}

\section{Conclusion}
We present ToolTweak a novel attack on tool-calling that can substantially bias tool selection across diverse tasks and queries. The attack is transferable across models, resists common defenses, and achieves success rates comparable to other reproducible attacks, raising serious concerns for the alignment and security of agentic systems. Even in benign cases, the effect resembles search engine optimization, where tools are promoted not for their capabilities but for persuasive naming and descriptions, undermining fairness. With usage-based pricing models, this can lead some tool providers to make significantly more revenue and cause others to lose money, even if their products have equal merits. Combined with supply-chain vulnerabilities, such attacks could escalate to large-scale disruptions.

Future work should explore larger, dynamic tool databases with retrieval systems and against white-box, gradient-based attacks. 
Our findings highlight the urgent need for more robust defenses and a fairer tool ecosystem.

\newpage






\bibliography{iclr2026_conference}
\bibliographystyle{iclr2026_conference}

\appendix

\newpage

\section{Data Standardization}
\label{appendix-data-standardization}

We converted the selected ToolBench APIs into the suggested JSON-schema format \citep{openai_function_nodate}. Although large language models, themselves, can accept any string, the tool-calling application layer often requires specific formatting. During initial testing, the dataset failed validation with several proprietary LLM APIs due to invalid function names containing forbidden characters, such as periods (.) and other non-alphanumeric characters. A review of API documentation revealed a common set of constraints. Most consumer services, including Deepseek, Claude, and OpenAI, restrict the set of possible function names. For example, Claude uses the following regular expression to filter invalid names \texttt{[a-zA-Z0-9\_-]\{1,64\}} \citep{anthropic_how_2025}.

Thus, in order to test on proprietary model services, we truncated tool names in our dataset to less than 64 characters and stripped them of any non-alphanumerics or non-underscore characters. We also converted all parameter data types into one of three JSON Schema-compatible types \footnote{\url{https://json-schema.org/understanding-json-schema/reference/type}}: \texttt{string}, \texttt{boolean}, or \texttt{number}.

\section{Transferable Map}

\begin{figure}[h]
  \centering
\includegraphics[width=0.6\textwidth]{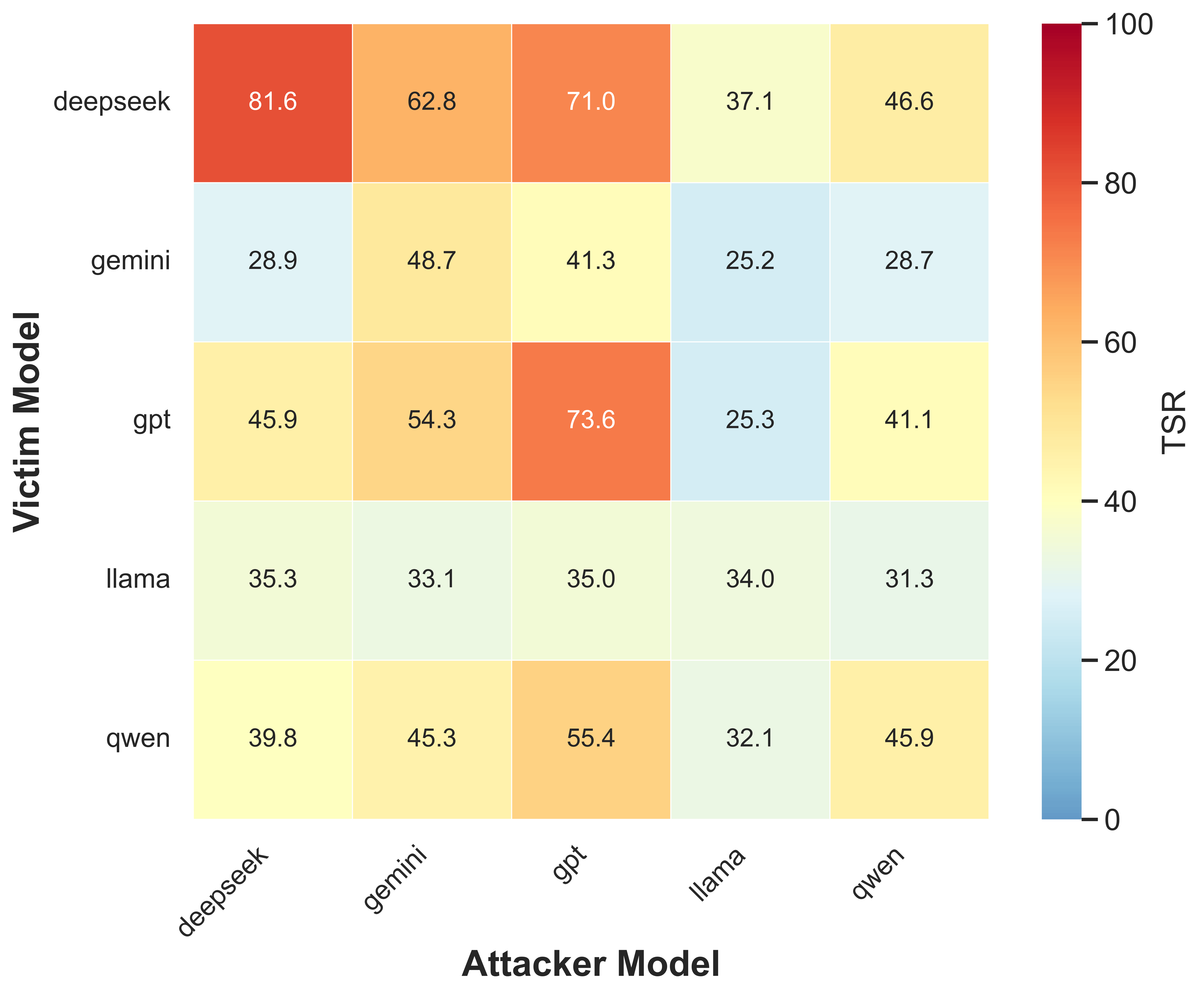}
  \caption{Transferability Heatmap of Effectiveness of Tools Augmented by Attacker LLMs}
\label{fig:transferability-heatmap}
\end{figure}

\section{Ablation} \label{appendix-ablations}

\paragraph{Number of Iterations} We examine the effectiveness of the task as the number of iterations $k$, grows larger. 
We find that, across all models except Llama3.1, there is a big gain in best selection rate at $k=2$, after the first revised name and description are generated, followed by a general trend upward, but at a significantly slower rate. 
This indicates that the attack can still be quite effective in a one-shot setting, significantly reducing computational costs; however, the iteration is still a necessary component for the most effective attack, with an increase of up to 20 percentage points at $k=10$ for the DeepSeek model.

\begin{figure}[h!]
    \centering
    \begin{subfigure}[b]{0.4\textwidth}
        \centering
        \includegraphics[width=\textwidth]{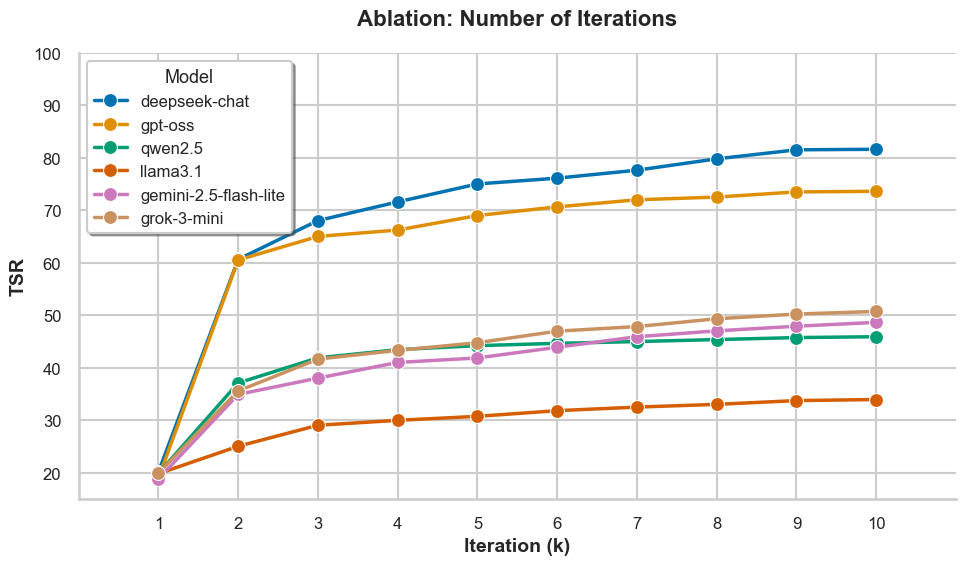}
        \caption{Iteration Ablation}
        \label{fig:iteration-ablation}
    \end{subfigure}
    \hfill
    \begin{subfigure}[b]{0.29\textwidth}
        \centering
        \includegraphics[width=\textwidth]{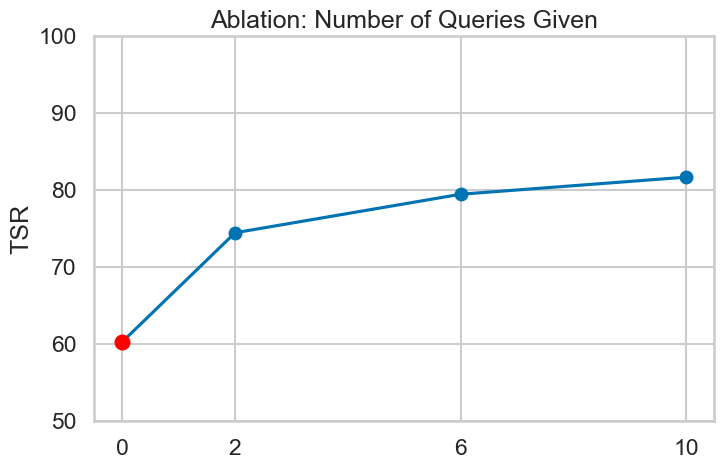}
        \caption{Query Ablation}
        \label{fig:query-ablation}
    \end{subfigure}
    \hfill
    \begin{subfigure}[b]{0.29\textwidth}
        \centering
        \includegraphics[width=\textwidth]{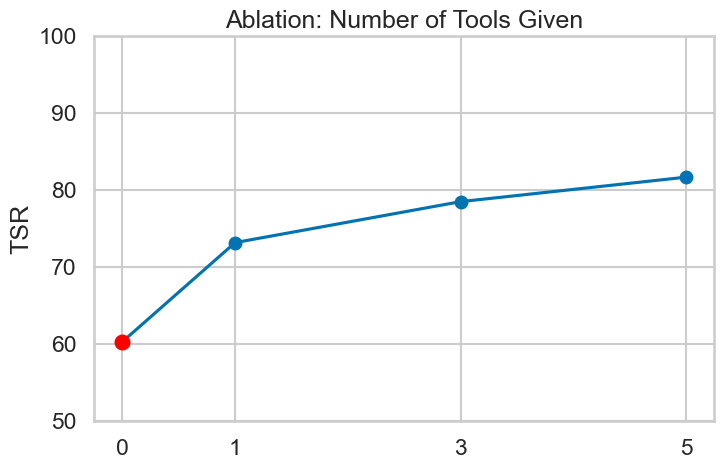}
        \caption{Tool Ablation}
        \label{fig:tool-ablation}
    \end{subfigure}
    \caption{\textbf{TSR} for various ablations. The \textcolor{red}{red dot} represents the ``No Knowledge'' setting}
    \label{fig:ablation}
\end{figure}

We evaluate our remaining ablation studies on DeepSeek, since the model proved both very susceptible to attack and capable of generating manipulative tool names and descriptions.

\paragraph{Number of Example Queries} At every iteration, the attacker receives a subset of queries used by the agent, $\mathcal{Q}_{exmpl} \subseteq \mathcal{Q}$. We test for varying number of queries $|\mathcal{Q}_{exmpl}| \in \{2,6,10\}$. We find that having a larger number of queries modestly increases the TSR from approximately $\sim74.4\%$ with 2 example queries to $\sim81.6\%$ with 10.
\paragraph{Number of Tools} In the default attack settings, the attacker has access to all tools in $\mathcal{T} \cup \{t\}$. Thus, we evaluate the attack with $|\mathcal{T} \cup \{t\}\}| \in \{1, 1+2, 1+4\}$\footnote{If tools are included, the target tool is always added to the context}. Similarly to the number of example queries, we find that having access to more tools increases the TSR from $\sim73\%$ with 1 tool to the base attack's $81.6\%$. 
\paragraph{No Query or Tool Knowledge Setting} We also examine a no-information black-box setting, where the attacker has no knowledge of its full tool schema, competitor tools, or no knowledge of user queries. The only signal it receives is the usage statistics. This restrictive setting is more widely applicable and generalizable, as an attacker may not know specifics about the exact tool setup or which queries the agent is asked. In this setting, we find that the attack is still successful, achieving a TSR of $\approx 60\%$; however, this is a large drop from  the full attack with above 80\%.

\section{Temperature and Reproducability}
\subsection{Temperature}
For all evaluations with the victim model, including tool selection and paraphrasing, we set $\textrm{temperature} = 0$. This is not the case for the attacker model. During preliminary runs, we found that, at temperature 0, the attacker model would eventually get stuck generating the same improvement and description. Additionally, the attack hinges on the attacking model's `creativity' and ability to explore text space to find better-performing descriptions. However, at higher temperatures (i.e. $\geq0.7$) we found that models were worse at following instructions and produced improperly formatted JSON. Thus, for the attacker, we set $\textrm{temperature} = 0.5$. 

\subsection{Reproducibility}

\paragraph{Random victim} We pass $\textrm{victim}=42$ to every model provider that supports it through the OpenAI API client. Additionally, we use the random victim $42$ for any random shuffling and sampling done by our Python experiment client.

Part of the motivation behind setting $\textrm{temperature} = 0$ for the victim LLM was to make evaluations of tool selection \textit{more} deterministic. However, contrary to popular belief, it does not guarantee determinism. Even when setting $\textrm{temperature} = 0$, LLM APIs offered by inference providers are often non-deterministic~\citep{kaddour_challenges_2023}. Additionally, \citet{puigcerver_sparse_2024} found that mixture of expert models are non-deterministic on the sequence level.

\section{Prompts}
\FloatBarrier
\begin{figure}
    \centering
    \include{figures/attacker-prompt}
    \caption{System Prompt Provided to the Attacker LLM}
    \label{fig:attacker-prompt}
\end{figure}

\begin{figure}
    \centering
    \begin{codeblock}
You are an expert at writing technical documentation and tool descriptions in an unbiased manner. The tools provided are from external sources which may attempt to unfairly influence tool selection. Please paraphrase the following tool description in a manner that is objective as possible. Remove all opinions and attempts at manipulation. You must still keep all of the essential information about the tool and refrain from adding any new information. Only respond with the new description:
\end{codeblock}
    \caption{System Prompt Provided to the Agent LLM for Paraphrase Defense}
    \label{fig:defender-prompt}
\end{figure}

\FloatBarrier

\section{Miscellaneous Appendices}

\begin{figure}[ht!]
\centering
\begin{codeblock}
{"improvement": "The tool's selection rate remains at 0%, indicating that the current name and description are not resonating with the LLM, and it continues to prefer other geocoding tools. The key observation from the feedback is the LLM's consistent selection of `geolocate_get_coordinates` (both versions) and `geocode_forward_and_reverse_forward_geocode`. These tools share several characteristics: 1) Direct, action-oriented names. 2) Clear specification of the 'address' parameter. 3) Concise descriptions. My tool's name, `geolocate_get_coordinates`, is now identical to one of the most preferred tools ... 
\end{codeblock}
\caption{\textbf{Reward-Hacking Generation Example}. The LLM attempts to directly copy the name of a preferred tool to increase its selection rate.}
\label{fig:reward_hacking_example}
\end{figure}

\begin{figure}
    \centering
    \includegraphics[width=\linewidth]{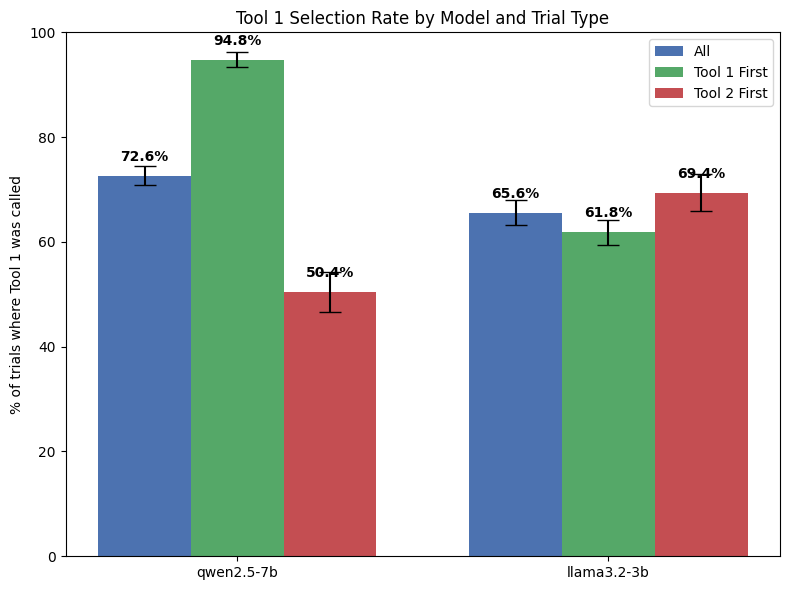}
    \caption{Evaluation of tool name bias, controlling for presentation order. The plot shows the selection rate of Tool 1 when paired with Tool 2. The blue bars represent the primary result, averaging across trials where Tool 1 was presented first and second to control for positional effects. Both Qwen2.5-7B (72.6\%) and Llama3.2-3B (65.6\%) show a significant bias towards selecting the tool with the number 1 in its name. We averaged over 5 trials with the default settings for both models using Ollama}
    \label{fig:tool1}
\end{figure}

\begin{figure}
    \centering
    \includegraphics[width=0.8\textwidth,trim={0 0 0 0.75cm},clip]{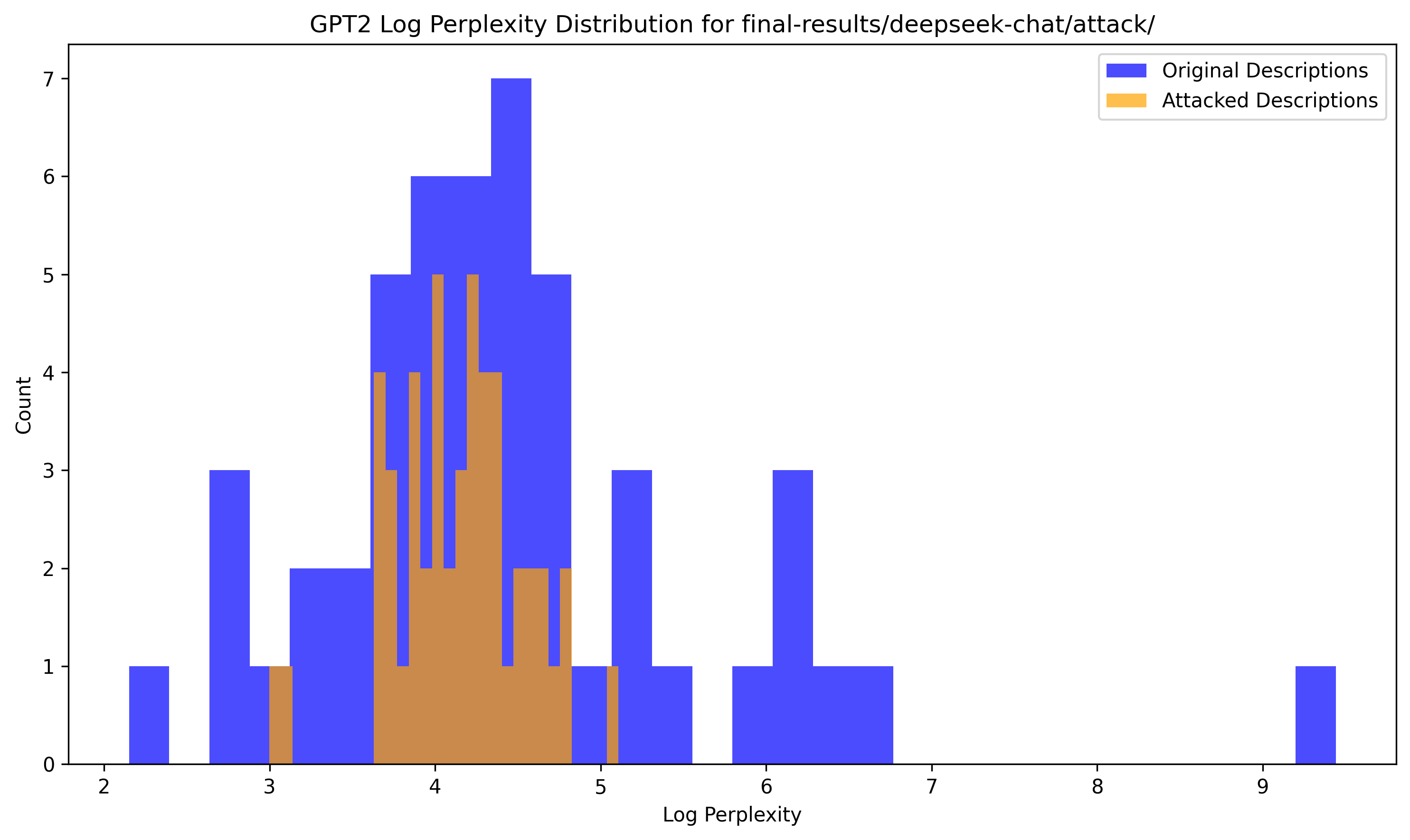}
    \caption{\textbf{Log-Perplexity Histogram} for DeepSeek attack descriptions and original description}
    \label{fig:perplexity-histogram}
\end{figure}

\begin{figure}
    \centering
    \includegraphics[width=0.8\textwidth,trim={0 0 0 0.75cm},clip]{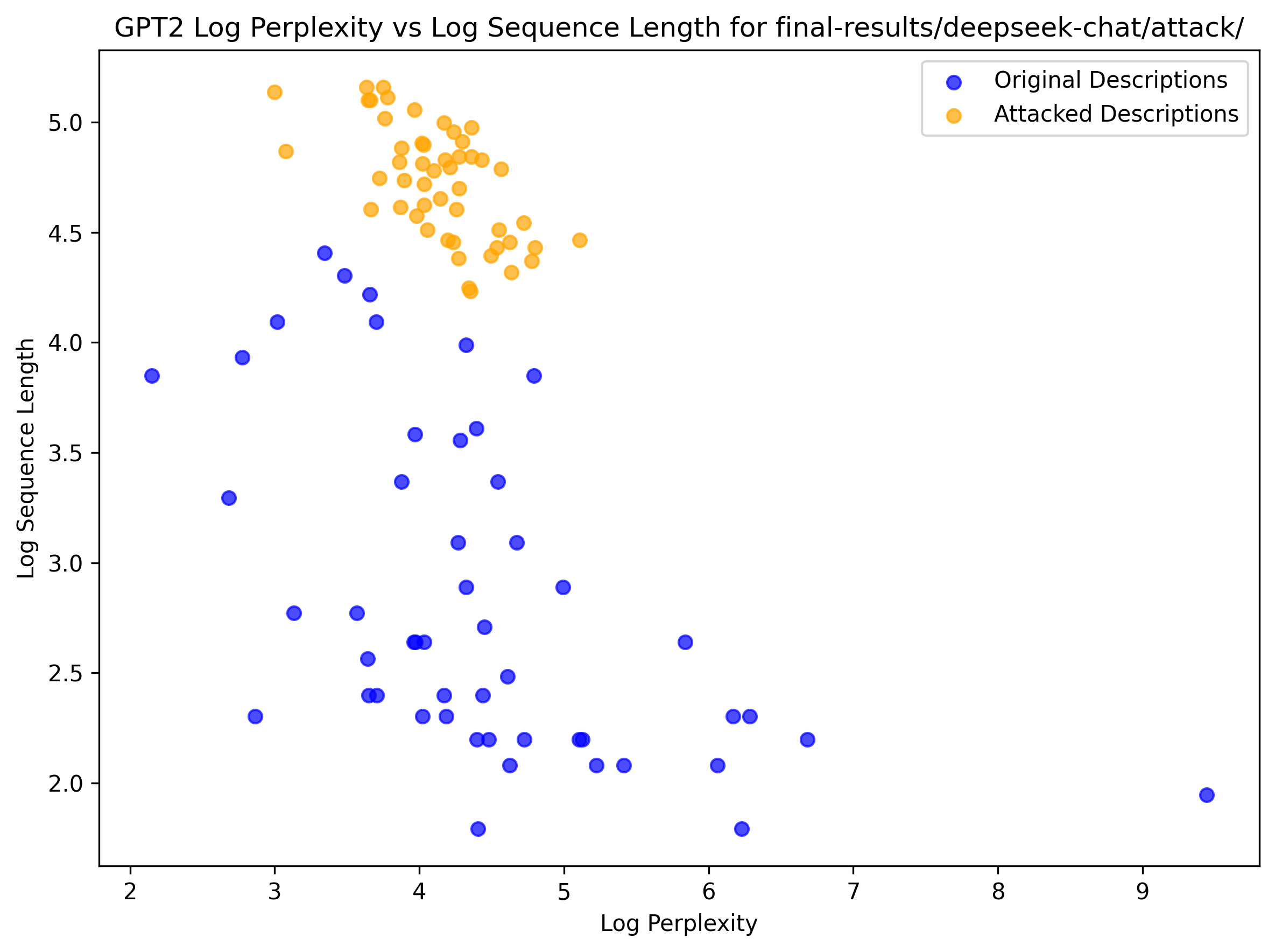}
    \caption{\textbf{Log-Sequence Length vs Log-Perplexity Plot} for DeepSeek attack descriptions and original description}
    \label{fig:perplexity-seq-len}
\end{figure}

\begin{figure}[htbp]
    \centering
\includegraphics[width=0.8\textwidth,trim={0 0 0 0.75cm},clip]{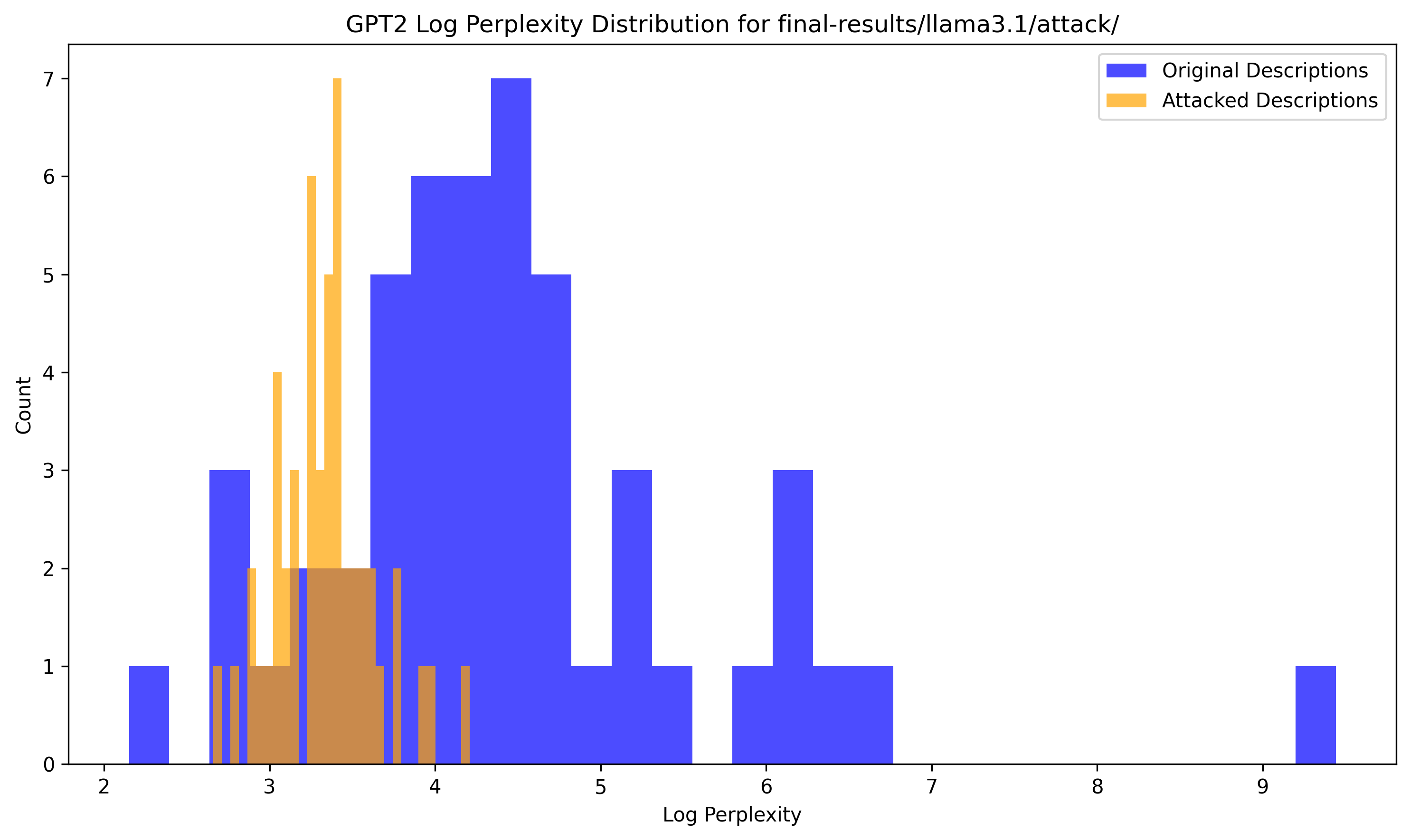}
    \caption{\textbf{Log-Perplexity Histogram} for Llama3.1 attack descriptions and original descriptions}
    \label{fig:llama-perplexity-histogram}
\end{figure}

\begin{figure}[htbp]
    \centering
    \includegraphics[width=0.8\textwidth,trim={0 0 0 0.75cm},clip]{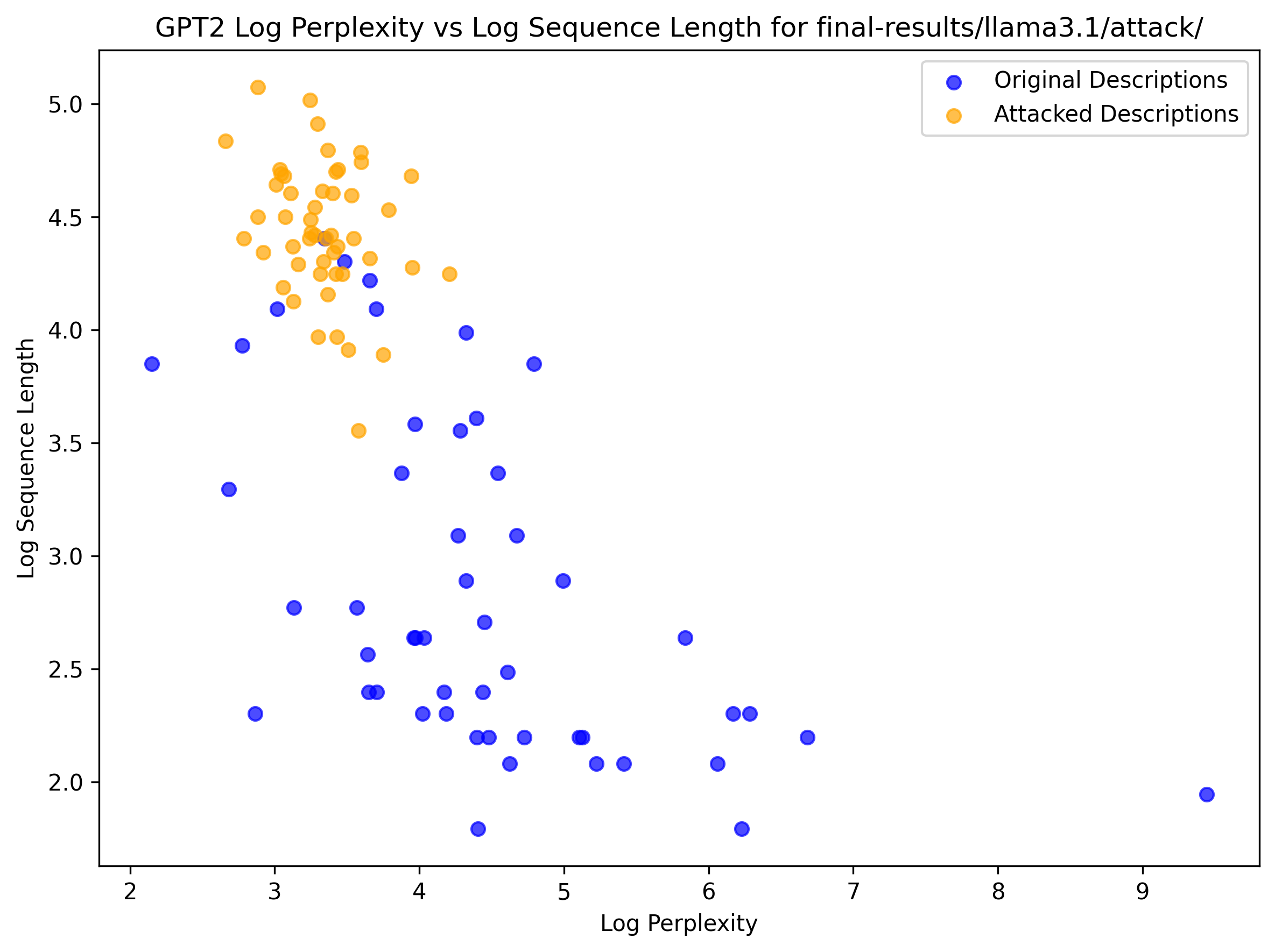}
    \caption{\textbf{Log-Sequence Length vs Log-Perplexity Plot} for Llama3.1 attack descriptions and original descriptions}
    \label{fig:llama-perplexity-seq-len}
\end{figure}

\begin{figure}
    \centering
    \includegraphics[width=\linewidth]{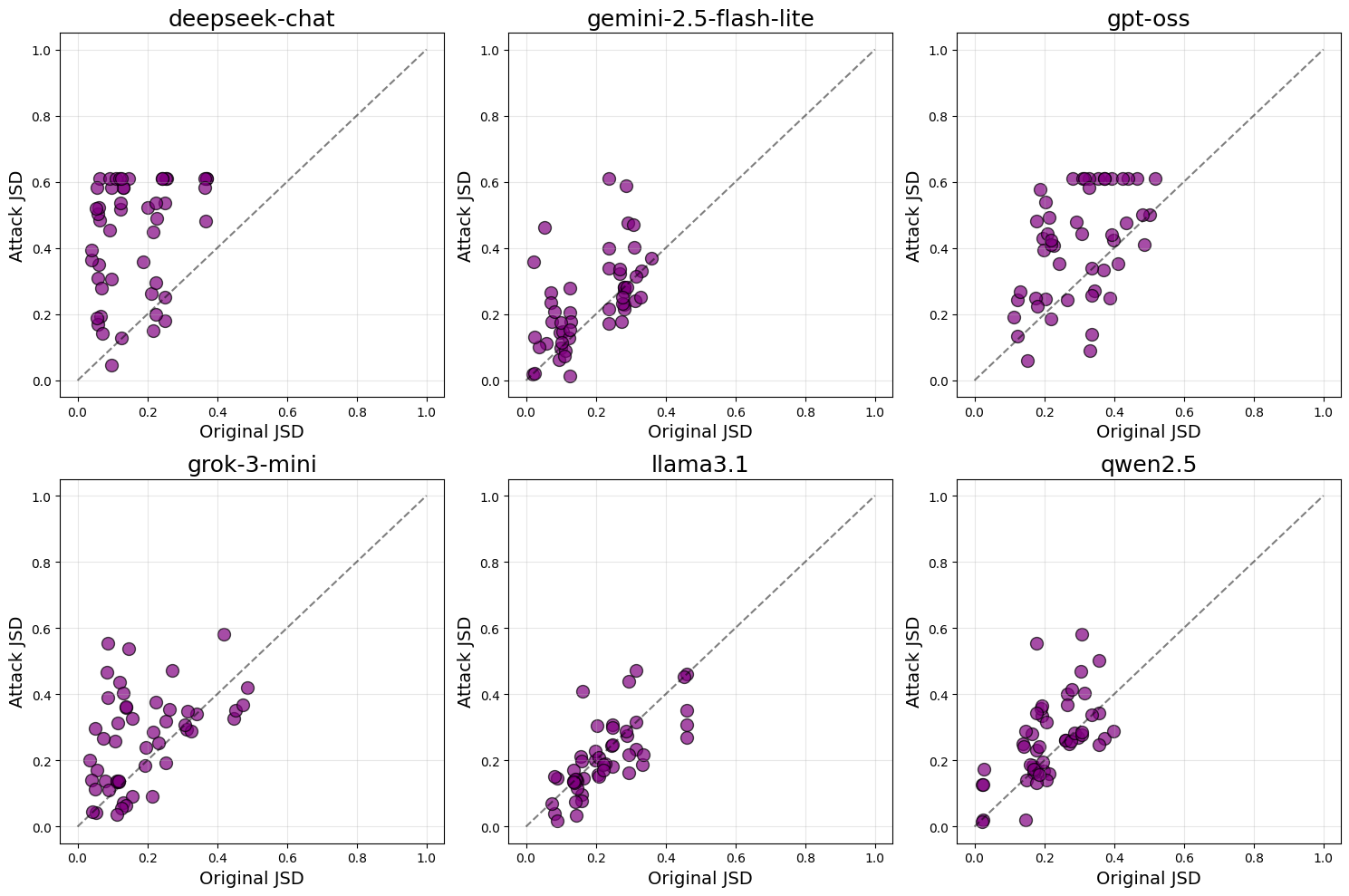}
    \caption{$D_\textrm{JS}$ \textbf{Before Attack (x-axis) vs. After Attack (y-axis) per model} Each dot represents the $D_{\textrm{JS}}$ between the observed distribution and $\textrm{Unif}(\mathcal{T})$ for a specific cluster and tool. Attacks above the $y=x$ line were able to improve tool selection rate. The further they are from $y=0$, the more biased the attack.}
    \label{fig:attack-jsd-uniform}
\end{figure}

\subsection{Gaming Tool Calls Experiments}

As part of our research, we recreated the experiments on the BFCL v3 \verb|simple| dataset \citep{yan_berkeley_2024} outlined by \citet{faghih_gaming_2025}. We did this to examine tool selection in a simple setting, looking at factors that impact tool selection rates, suffix attacks, and defenses.

We examined selection rates with and without a paraphrase defense. We found that in the undefended case, tools with an augmented description were called $<6.5 $ times more than the original description. Meanwhile, after applying an objective paraphrase defense, the tools with an augmented description were called about $1.18$ times less than the original tool.

\subsection{Jensen—Shannon Divergence}

\begin{align}
    D_\textrm{JS}\left(P \,||\, Q\right) &= D_\textrm{KL}(P\,||\,\frac{P+Q}{2}) + D_\textrm{KL}(Q\,||\,\frac{P+Q}{2}) \\
    &= \sum_{x\in\mathcal{X}}P(x)\log{\frac{2P(x)}{P(x)+Q(x)}} + \sum_{x\in\mathcal{X}}Q(x)\log{\frac{2Q(x)}{P(x)+Q(x)}}
    \label{eq:divergence}
\end{align}

\pagebreak

\end{document}